\documentclass[a4paper,aps,11pt,nofootinbib,pdftex]{revtex4}
\usepackage{amsmath}
\usepackage{amssymb}
\usepackage{amsfonts}
\usepackage{ascmac}
\usepackage{color}
\usepackage{setspace}

\usepackage{graphicx}
\begin{document}

%%%%%%%%%%%%%%%%%%%%%%%%%%%%%%%%%%%%%%%
\title{%\large
	Homogeneous Balls in a Spontaneously Broken U(1) Gauge Theory
}
\vspace{2cm}

%\today
\begin{spacing}{1.0}
\hfill{OCU-PHYS 496}
\\
\hfill{AP-GR 153}
\\
\hfill{NITEP 7}
\end{spacing}

\vspace{2cm}

%%%%%%%%%%%%%%%%%%%%%%%%%%%%%%%%%%%
\author{\vspace{1cm}\large
Hideki Ishihara}
\email{ishihara@sci.osaka-cu.ac.jp}
\author{Tatsuya Ogawa}
\email{taogawa@sci.osaka-cu.ac.jp}
\affiliation{\vspace{0.5cm}
Department of Mathematics and Physics, Graduate School of Science,
	\\ and \\
Nambu Yoichiro Institute of Theoretical and Experimental Physics (NITEP),
Osaka City University, Osaka 558-8585, Japan}
%%%%%%%%%%%%%%%%%%%%%%%%%%%%%%%%%%%%%%%%

%%%%%%%%%%%%%%%%%%%%%%%%%%%%%%%%%%%%%%%%
\begin{abstract}
\vspace{0.5cm}

We study the coupled system consisting of a complex matter scalar field, a U(1) gauge field,
and a complex Higgs scalar field that causes spontaneously symmetry breaking.
We show by numerical calculations that there are spherically symmetric nontopological
soliton solutions.
Homogeneous balls solutions, all fields take constant values inside the ball and
in the vacuum state outside, appear in this system.
It is shown that the homogeneous balls have the following properties:
charge density of the matter scalar field is screened
by counter charge cloud of the Higgs and gauge field everywhere;
an arbitrary large size is allowed;
energy density and pressure of the ball behave homogeneous nonrelativistic gas;
a large ball is stable against dispersion into free particles and against decay
into two smaller balls.
\end{abstract}
%%%%%%%%%%%%%%%%%%%%%%%%%%%%%%%%%%%%%%%%

\maketitle

%\tableofcontents

\newpage
%%%%%%%%%%%%%%%%%%%%%%%%%%%%%%%%%%%%%%%%%%%%%%%%%%%%%%%%%%%%%
\section{Introduction}
%%%%%%%%%%%%%%%%%%%%%%%%%%%%%%%%%%%%%%%%%%%%%%%%%%%%%%%%%%%%%

A class of interesting excitations in field theories is solitons, i.e.,
nonlinear solutions localized in finite spatial regions.
The solitons are classified into two types: topological and nontopological solitons.
The former are field configurations with topological charge that is invariant under
continuous deformations of the field with fixed boundary conditions.
The topological solitons cannot relax to zero energy configurations due to conserved
topological quantities.
The latter represent field configurations with the lowest energy for fixed conserved charge
in global U(1)-invariant theories, where the symmetry of the systems guarantee the stability.
Friedberg, Lee and Sirlin \cite{Friedberg:1976me} introduced nontopological solitons
in a coupled system of a complex scalar field and a self-interacting real scalar field.
Coleman \cite{Coleman:1985ki} showed the simplest example of nontopological solitons,
so-called Q-balls\footnote{
Hereafter, we call a spherically symmetric nontopological soliton a Q-ball, in short.},
can appear in a system of a self-interacting single complex scalar field.

The Q-balls attract much attention because the Q-balls generally appear
in theories with potentials inspired by supersymmetric theories
that include global U(1) symmetries \cite{Kusenko:1997zq, Dvali:1997qv, Kasuya:2000sc}.
Furthermore, in a cosmological context, the Q-ball is a candidate of the dark matter of
the universe \cite{Kusenko:1997si, %Kawasaki:2008zz,
Kusenko:2001vu, %Multamaki:2002hv,
Fujii:2001xp, Enqvist:2001jd, Kusenko:2004yw}
and a source for baryogenesys \cite{Enqvist:1997si, Kasuya:1999wu, Kawasaki:2002hq}.

Generalizations of the Q-balls
in local U(1)-invariant theories by introduction of a gauge field
are also studied \cite{Lee:1988ag,Shi:1991gh, Arodz:2008nm, Tamaki:2014oha,
Gulamov:2015fya}.
There are significant differences between gauged and ungauged Q-balls.
For example, an ungauged Q-ball with arbitrary large charge is
allowed while upper bound of charge appears
for a gauged Q-ball \cite{Lee:1988ag,Shi:1991gh, Gulamov:2015fya}.
Otherwise, complicated form of potential should be assumed for existence of
large Q-balls \cite{Arodz:2008nm, Tamaki:2014oha}.

In this paper, we consider a gauge theory with spontaneous symmetry breaking,
which is a fundamental framework of modern physics.
We study the system consisting of a complex scalar field as matter, a U(1) gauge field,
and a complex Higgs scalar field that causes spontaneous symmetry breaking:
a local U(1) $\times$ global U(1) symmetry breaks to a global U(1) symmetry.
While models with a single scalar field are assumed to have complicated self-interactions, e.g.,
third or sixth order potentials, or non-polynomial potentials,
for the existence of Q-balls,
we show the existence of Q-balls in the model that
has simple natural interaction terms.
Then, this work would suggest Q-balls can appear in a wide class of gauge theories.

We assume stationary and spherically symmetric configurations of the fields, and reduce
the system into a coupled ordinary differential equations.
We show Q-balls exist in this theory by using numerical method\footnote{
This was reported briefly in ref.\cite{Ishihara:2018rxg}.}.
The all fields are nonvanishing in a finite region while the matter scalar field and the gauge field
vanish, and the Higgs field takes the vacuum expectation value outside the region.
Phase rotation of the complex Higgs scalar field is absorbed by the gauge field,
and phase rotation of the complex matter scalar field, which represents charge, characterizes the
solutions.
There are two types of solutions classified by the shape: Gaussian balls, expressed by the Gaussian-like
functions, and homogeneous balls, expressed by step-like functions.
In this paper, we concentrate on the homogeneous balls, which are described by bounce solutions
that connect two stationary points of the ordinary differential equations, and clarify their properties.

We show that the homogeneous balls in the present system have the following properties.
The charge density of the matter scalar field of a homogeneous ball
is screened everywhere by a counter charge cloud of the Higgs and gauge fields, namely,
perfect screening occurs \cite{Ishihara:2018eah}.
A homogeneous ball has constant energy density and pressure inside the ball,
and the pressure is much smaller than the energy density, i.e., the homogeneous ball is like a
ball of non-relativistic gas.
A homogeneous ball with arbitrary large size is allowed in contrast to a gauged Q-ball without Higgs field
has upper limit of size.
A large homogeneous ball is stable against dispersion into free particles and
decay into two smaller Q-balls.

The paper is organized as follows. In Section II we present the basic model investigated in this paper,
and show that the model is described by a coupled system of ordinary differential equations.
In Section III, we obtain numerical solutions that represent Q-balls, and we see charge screening.
We analyze properties of the homogeneous ball solutions in Section IV
and stability of Q-balls in the present system in Section V.
Section VI is devoted to summary and discussions.

%%%%%%%%%%%%%%%%%%%%%%%%%%
\section{Basic Model}
%%%%%%%%%%%%%%%%%%%%%%%%%%
We consider the action given by
\begin{align}
	S=\int d^4 x \left(
  		-(D_{\mu}\psi)^*(D^{\mu}\psi) -(D_{\mu}\phi)^*(D^{\mu}\phi)
  		-V(\phi)-\mu \psi^{\ast} \psi \phi^{\ast}\phi -\frac{1}{4}F_{\mu\nu}F^{\mu\nu}
\right),
\label{eq:action}
 \end{align}
where $\psi$ is a complex matter scalar field, $\phi$ is a complex Higgs scalar field with the potential
\begin{align}
 	V(\phi):=\frac{\lambda}{4}(\phi^{\ast}\phi-\eta^2)^2,
  \label{eq:potential}
\end{align}
where $\lambda$ and $\eta$ are positive constants, and
$F_{\mu\nu}:=\partial_{\mu}A_{\nu}-\partial_{\nu}A_{\mu}$ is the field strength of
a U(1) gauge field $A_{\mu}$.
The covariant derivative $D_{\mu}$ in \eqref{eq:action} is defined by
\begin{align}
  	D_{\mu}\psi :=\partial_{\mu}\psi -ieA_{\mu}\psi,  \quad
	D_{\mu}\phi :=\partial_{\mu}\phi -ieA_{\mu}\phi,
 \label{eq:covariant_derivative}
\end{align}
where $e$ is a gauge coupling constant.
This model is a generalization of the Friedberg-Lee-Sirlin model
by introducing a complex Higgs scalar field and a $\text{U}(1)$ gauge field.

The action (\ref{eq:action}) is invariant under the local $\text{U}(1)$ times the global $\text{U}(1)$ gauge
transformations,
\begin{align}
  	&\psi(x) \to \psi'(x)=e^{i(\chi(x)-\gamma)}\psi(x),
 \label{eq:psi_tr} \\
  	&\phi(x) \to \phi'(x)=e^{i(\chi(x)+\gamma)}\phi(x),
 \label{eq:phi_tr} \\
 	&A_{\mu}(x)\to A_{\mu}'(x)=A_{\mu}(x)+e^{-1}\partial_{\mu}\chi(x),
 \label{eq:A_tr}
\end{align}
where $\chi(x)$ and $\gamma$ are an arbitrary function and a constant, respectively.
Owing to the gauge invariance, there are the conserved current
\begin{align}
  j_\psi^{\nu} &:=ie\left\{\psi^{\ast}(D^{\nu}\psi)-\psi(D^{\nu}\psi)^{\ast}\right\},
  \label{eq:j_psi}
  \\
  j_\phi^{\nu} &:=ie\left\{\phi^{\ast}(D^{\nu}\phi)-\phi(D^{\nu}\phi)^{\ast}\right\}
  \label{eq:j_phi}
\end{align}
satisfying $\partial_{\mu}j_{\psi}^{\mu}$=0 and $\partial_{\mu}j_{\phi}^{\mu}$=0.
Consequently, the total charge of $\psi$ and $\phi$ defined by
\begin{align}
  Q_\psi &:=\int \rho_{\psi} d^3x,
  \label{eq:Q_psi}
  \\
  Q_\phi &:=\int \rho_{\phi} d^3x,
  \label{eq:Q_phi}
\end{align}
are conserved, where $\rho_{\psi}:=j_{\psi}^t$ and $\rho_{\phi}:=j_{\phi}^t$.

The energy of the system is given by\footnote{
See Appendix \ref{E_M_tensor}.
}
\begin{align}
 	E=\int d^3x \biggl( \left|D_{t}\psi\right|^2 & +(D_{i}\psi)^{\ast}(D^{i}\psi)
		 +\left|D_{t}\phi\right|^2+(D_{i}\phi)^{\ast}(D^{i}\phi)\notag \\
	  	& +V(\phi)+\mu|\psi|^2|\phi|^2 +\frac{1}{2}\left(E_iE^i+B_iB^i\right)\biggl) ,
\label{eq:energy}
\end{align}
where $E_i:=F_{i0}$,  $B^i:=1/2\epsilon^{ijk}F_{jk}$, and $i$ denotes a spatial index.
In the vacuum state, which minimizes the energy \eqref{eq:energy}, the fields $\psi$,
$\phi$, and $A_{\mu}$ should satisfy
\begin{align}
  \psi=0,~ \phi^{\ast}\phi=\eta^2,~D_{\mu}\phi=0, ~ \text{and} ~ F_{\mu\nu}=0,
  \label{eq:VEV2}
\end{align}
equivalently,
\begin{align}
  \psi=0,~ \phi = \eta e^{i\theta(x)},~   \text{and}~  A_{\mu}=e^{-1}\partial_{\mu}\theta,
  \label{eq:VEV}
\end{align}
where $\theta$ is an arbitrary continuous regular function.
We exclude topologically non-trivial case in this paper.
The Higgs scalar field $\phi$ has the vacuum expectation value $\eta$,
then the $\text{U}_\text{local}(1)\times \text{U}_\text{global}(1)$ symmetry is broken
into a global $\text{U}(1)$ symmetry,
so that the gauge field $A_{\mu}$ and the complex scalar field $\psi$ acquire
the mass $m_A=\sqrt{2}e\eta$ and $m_{\psi}=\sqrt{\mu}\eta$, respectively.
The real scalar field that denotes a fluctuation of the amplitude of $\phi$
around $\eta$ acquires the mass $m_\phi=\sqrt{\lambda}\eta$.

By varying (\ref{eq:action}) with respect to $\psi^{\ast}$, $\phi^{\ast}$, and $A_{\mu}$,
we obtain the equations of motion
\begin{align}
  &D_{\mu}D^{\mu}\psi-\mu \phi^{\ast}\phi \psi =0,
  \label{eq:equation of psi}\\
 &D_{\mu}D^{\mu}\phi-\frac{\lambda}{2}\phi(\phi^{\ast}\phi-\eta^2)-\mu \phi \psi^{\ast}\psi =0,
  \label{eq:equation of phi}\\
 &\partial_{\mu}F^{\mu\nu}=j_\phi^{\nu}+j_{\psi}^{\nu}.
  \label{eq:equation of gauge field}
\end{align}

%------------------------------------------
%------------------------------------------
We assume that the fields are stationary and spherically symmetric
in the form,
\begin{align}
  	&\psi=e^{i\omega t}u(r),
  \label{eq:psi_ansatz}\\
  	&\phi=e^{i\omega' t}f(r),
  \label{eq:phi_ansatz}\\
 	&A_t=A_t(r),\quad \mbox{and}\quad A_i=0,
  \label{eq:A_ansatz}
\end{align}
where $\omega$ and $\omega'$ are constants, $u(r)$ and $f(r)$ are real functions of $r$.
Using the gauge transformation \eqref{eq:psi_tr}, \eqref{eq:phi_tr} and \eqref{eq:A_tr},
we fix the variables as
\begin{align}
  &\phi(r) \to f(r),
\label{eq:f}\\
  &\psi(t,r) \to e^{i\Omega t}u(r):=e^{i(\omega-\omega') t}u(r),
\label{eq:u}\\
  &A_t(r) \to \alpha(r):= A_t(r)-e^{-1}\omega',
\label{eq:alpha}
\end{align}
where we assume $\Omega:=\omega-\omega'>0$ without loss of generality.

Substituting \eqref{eq:f}, \eqref{eq:u}, and \eqref{eq:alpha}
into \eqref{eq:equation of psi}, \eqref{eq:equation of phi}, and \eqref{eq:equation of gauge field},
we obtain a set of the ordinary differential equations:
\begin{align}
 &\frac{d^2u}{dr^2}+\frac{2}{r}\frac{du}{dr}+(e\alpha-\Omega)^2u-\mu f^2u=0,
 \label{eq:eq_u}\\
 &\frac{d^2f}{dr^2}+\frac{2}{r}\frac{df}{dr}+e^2f \alpha^2-\frac{\lambda}{2}f(f^2-\eta^2)-\mu fu^2=0,
\label{eq:eq_f}\\
 &\frac{d^2\alpha}{dr^2}+\frac{2}{r}\frac{d\alpha}{dr}+\rho_\text{total}=0,
\label{eq:eq_alpha}
\end{align}
where $\rho_\text{total}$ is defined by
\begin{align}
  \rho_\text{total}(r):=\rho_{\psi}(r)+\rho_{\phi}(r).
  \label{eq:rho_total}
\end{align}
The charge densities $\rho_\psi$ and $\rho_\phi$ are given by the variables $u, f$, and $\alpha$ as
\begin{align}
	\rho_\psi &= -2e(e\alpha-\Omega) u^2 ,
\label{eq:rho_psi}
\\
	\rho_\phi &= -2e^2\alpha f^2.
\label{eq:rho_phi}
\end{align}
We seek configurations of the fields with a non-vanishing value of $\Omega$ that characterizes
the solutions.

We require boundary conditions for the fields so that the fields should be regular
at the origin.
Then, we impose the conditions for the spherically symmetric fields as
\begin{align}
  \frac{du}{dr}\to 0 \ , \
  \frac{df}{dr}\to 0 \ , \ \frac{d\alpha}{dr}\to 0 \quad \mbox{as}\quad r\to 0.
\label{eq:BC_origin}
\end{align}
On the other hand, fields should be in the vacuum state at the spatial infinity.
Therefore, from \eqref{eq:VEV} we impose the conditions
\begin{align}
  u \to 0 \ , \ f \to \eta \ , \ \alpha \to 0 \quad \mbox{as}\quad r\to \infty .
\label{eq:BC_infty}
\end{align}

%%%%%%%%%%%%%%%%%%%%%%%%%%%%%%%%%%%%%
\section{Numerical Calculations}
%%%%%%%%%%%%%%%%%%%%%%%%%%%%%%%%%%%%%
In this section, we present numerical solutions of the Q-ball by using the relaxation method.
In numerics, hereafter, we set $\eta$ as the unit,
and scale the radial coordinate $r$ as $r \to \eta r$,
and scale the functions $f$, $u$, $\alpha$ as $f\to \eta^{-1}f$, $u\to \eta^{-1}u$,
$\alpha \to \eta^{-1}\alpha$, respectively,
and scale the parameter $\Omega$ as $\Omega \to \eta^{-1}\Omega$.
We set $\lambda=1$, $\mu=1.4$ and $e=1$, as an example, in this paper.

\begin{figure}[!ht]
\centering
\includegraphics[width=7.3cm]{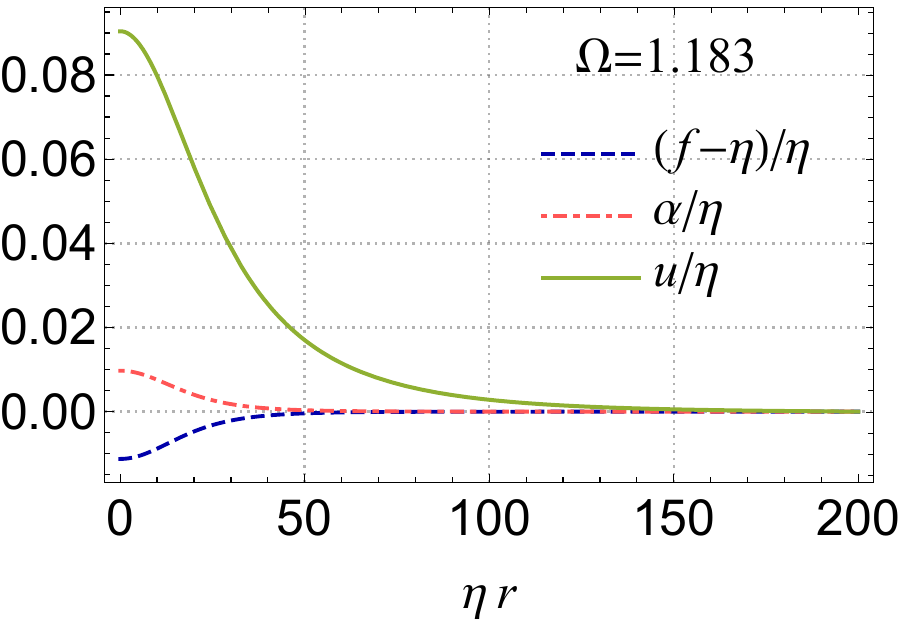}~~~~
\includegraphics[width=7.3cm]{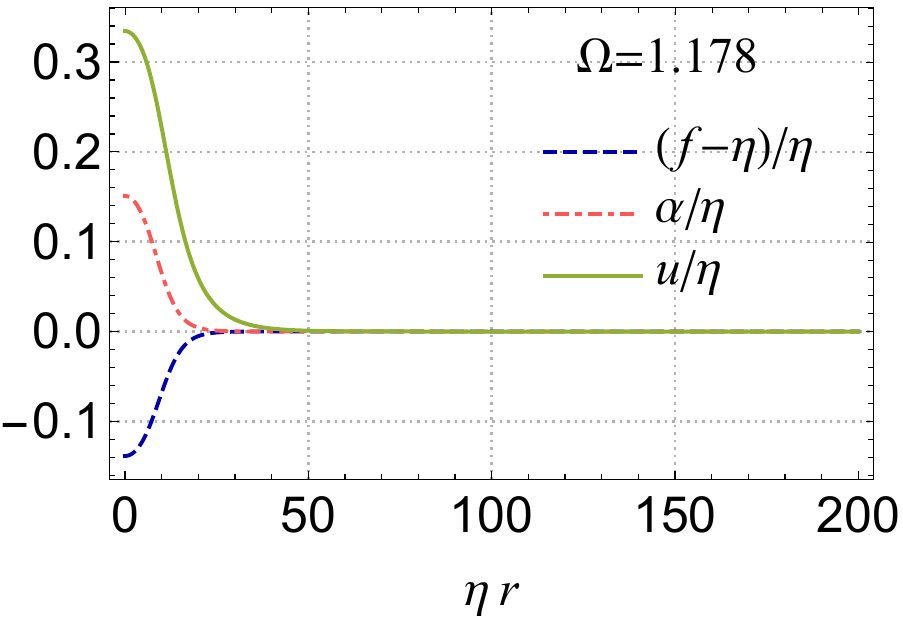}
\\
\includegraphics[width=7.3cm]{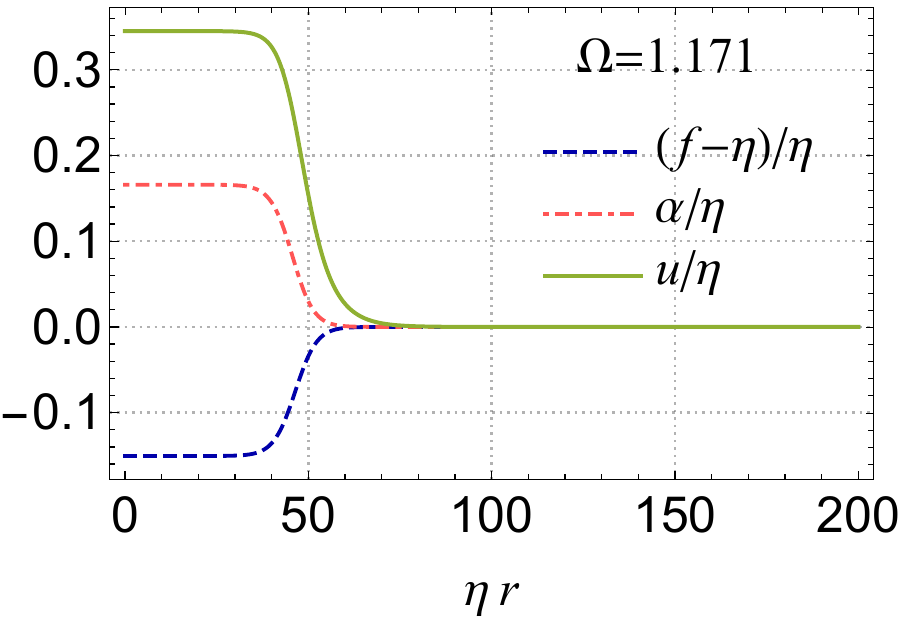}~~~~
\includegraphics[width=7.3cm]{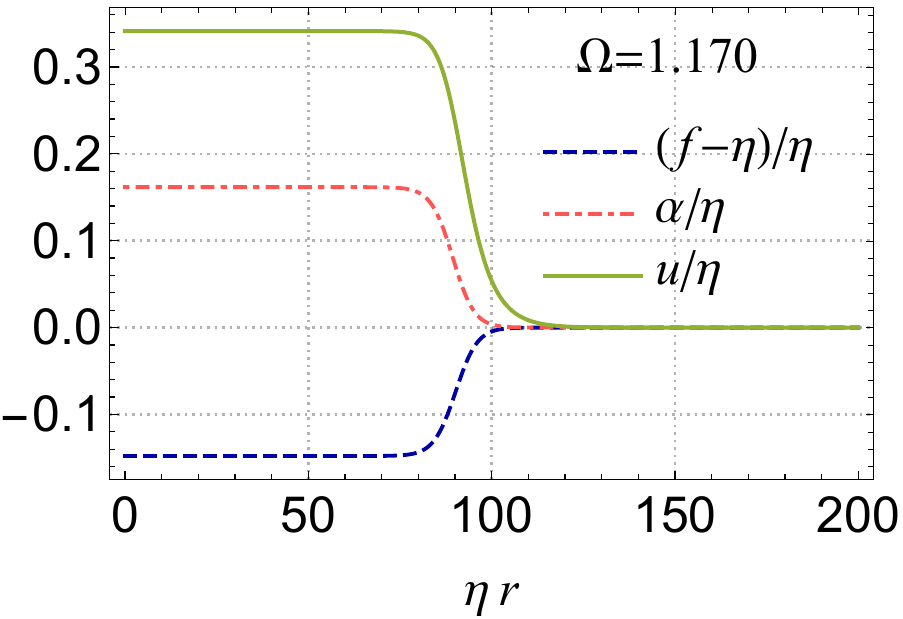}
\caption{
Numerical solutions $f(r)$, $u(r)$, and $\alpha(r)$ are drawn for $\Omega=1.183, 1.178, 1.171$ ,
and $1.170$.
\label{fig:configurations_1p4}
}
\end{figure}

%--------------------------------------------------
%--------------------------------------------------
In Fig.\ref{fig:configurations_1p4}, we plot $u(r)$, $f(r)$,
and $\alpha(r)$ as functions of $r$ with four values of $\Omega$.
In the all cases of $\Omega$, the functions, whose shapes depend on $\Omega$,
have finite support, namely, solitary solutions are obtained.

%%%%%%%%%%%%%%%%%%%%%%%%%%%%%%%%%%%%%%%%%%%%%%%%%%%%%%%%%%%%%%%%%%%%%%%%%%%%%%%%%%%%%%
%%%%%%%%%%%%%%%%%%%%%%%%%%%%%%%%%%%%%%%%%%%%%%%%%%%%%%%%%%%%%%%%%%%%%%%%%%%%%%%%%%%%%%

In the case of $\Omega=1.183$ and $\Omega=1.178$, the field profiles are Gaussian function like.
On the other hand, for $\Omega=1.171$, $\Omega=1.170$, the field profiles are step function like.
The solutions in the latter cases represent homogeneous balls, namely,
the functions $u, f$ and $\alpha$ take constant values inside the ball,
and they change the values quickly in a thin region of the ball surface, $r=r_s$,
and $u$, $\alpha$ vanish and $f$ takes the vacuum expectation value $\eta$ outside the ball.

%--------------------------------------
%--------------------------------------
%%%%%%%%%%%%%%%%%%%%%%%%%%%%%%%%%%%%%%%%%%%%%%%%%%%%%%%%%%%%%%%%%%%%%%%%%%%%%%%%%%%%%%
\begin{figure}[!ht]
\centering
\includegraphics[width=7.5cm]{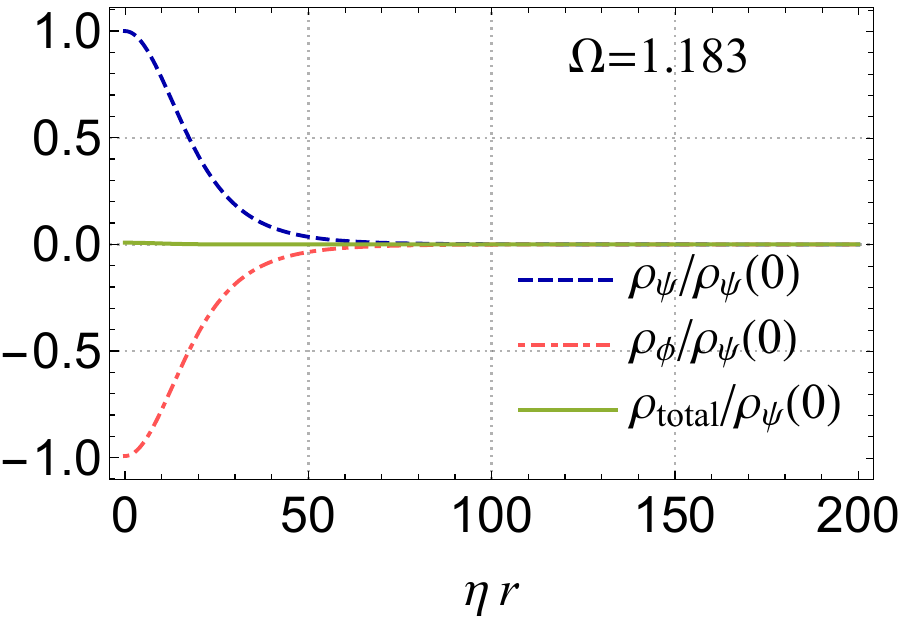}~~~~
\includegraphics[width=7.5cm]{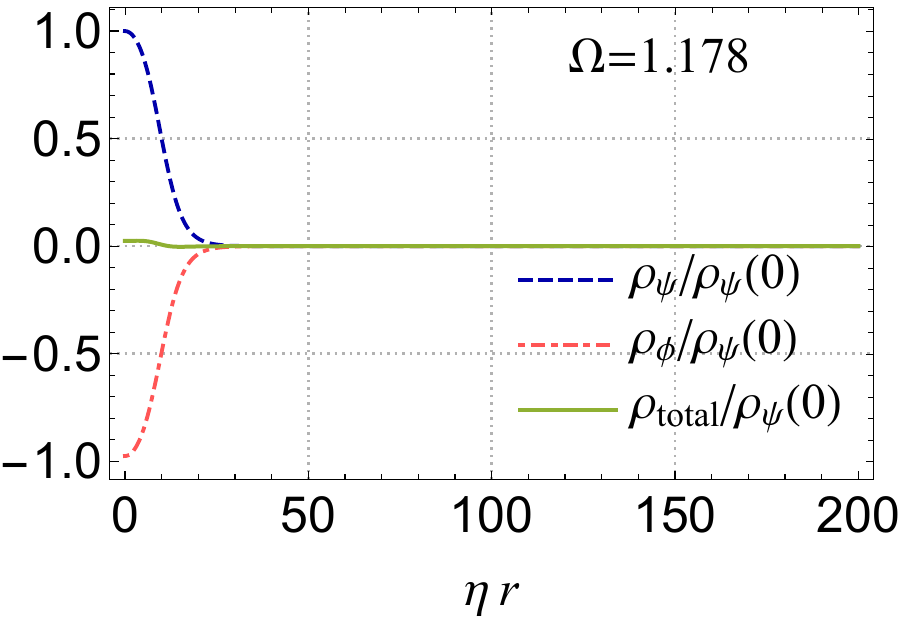}
\\
\includegraphics[width=7.5cm]{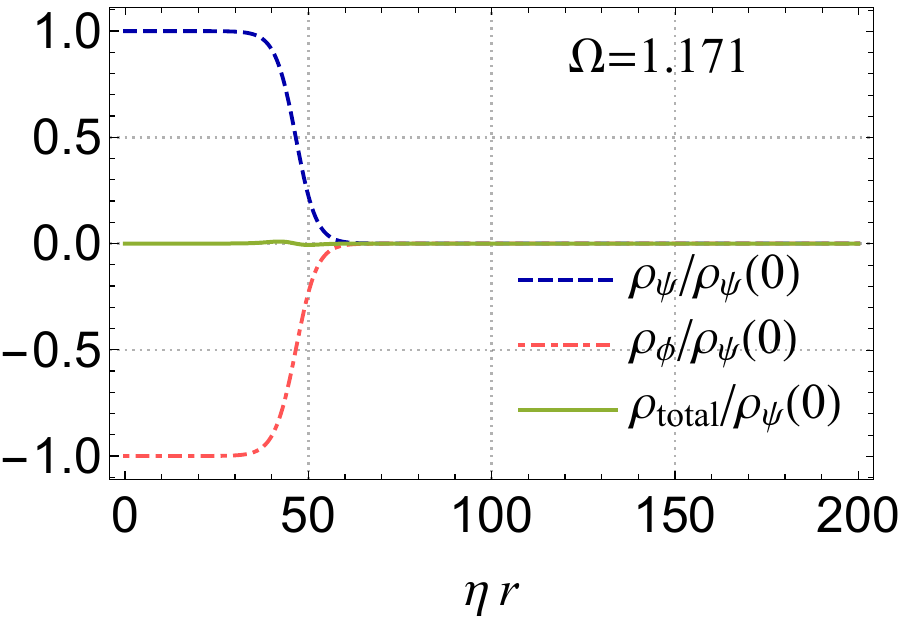}~~~~
\includegraphics[width=7.5cm]{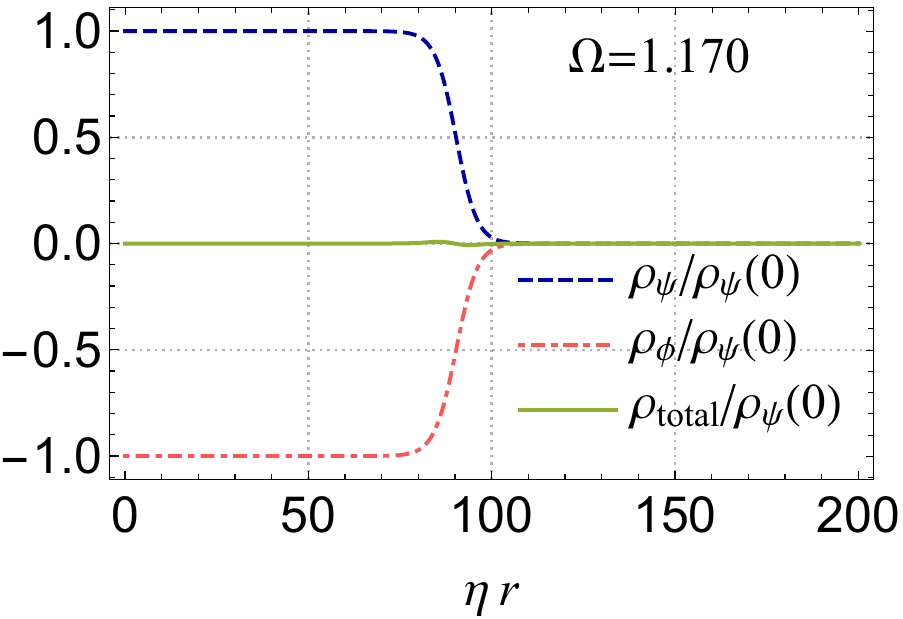}
\begin{minipage}{0.45\hsize}
\vspace{1cm}
\end{minipage}
\\
\caption{
The charge densities $\rho_{\psi}$, $\rho_{\phi}$ and
$\rho_\text{total}:=\rho_{\psi}+\rho_{\phi}$ normalized by the central value of $\rho_{\psi}$
are shown for $\Omega=1.183, 1.178, 1.171$, and $1.170$.
\label{fig:charge_mu1p4}
\vspace{1cm}
}
\end{figure}
%%%%%%%%%%%%%%%%%%%%%%%%%%%%%%%%%%%%%%%%%%%%%%%%%%%%%%%%%%%%%%%%%%%%%%%%%%%%%%%%%%%%%%

By numerical calculations, we depict the charge densities $\rho_{\psi}(r)$ and $\rho_{\phi}(r)$
in Fig.\ref{fig:charge_mu1p4} as functions of $r$.
We find that the charge density $\rho_\psi$ is compensated by the counter charge density $\rho_\phi$.
Then, $\rho_\text{total}$ almost vanishes everywhere, namely,
perfect screening occurs \cite{Ishihara:2018eah}.

\begin{figure}[!ht]
\centering
\includegraphics[width=8cm]{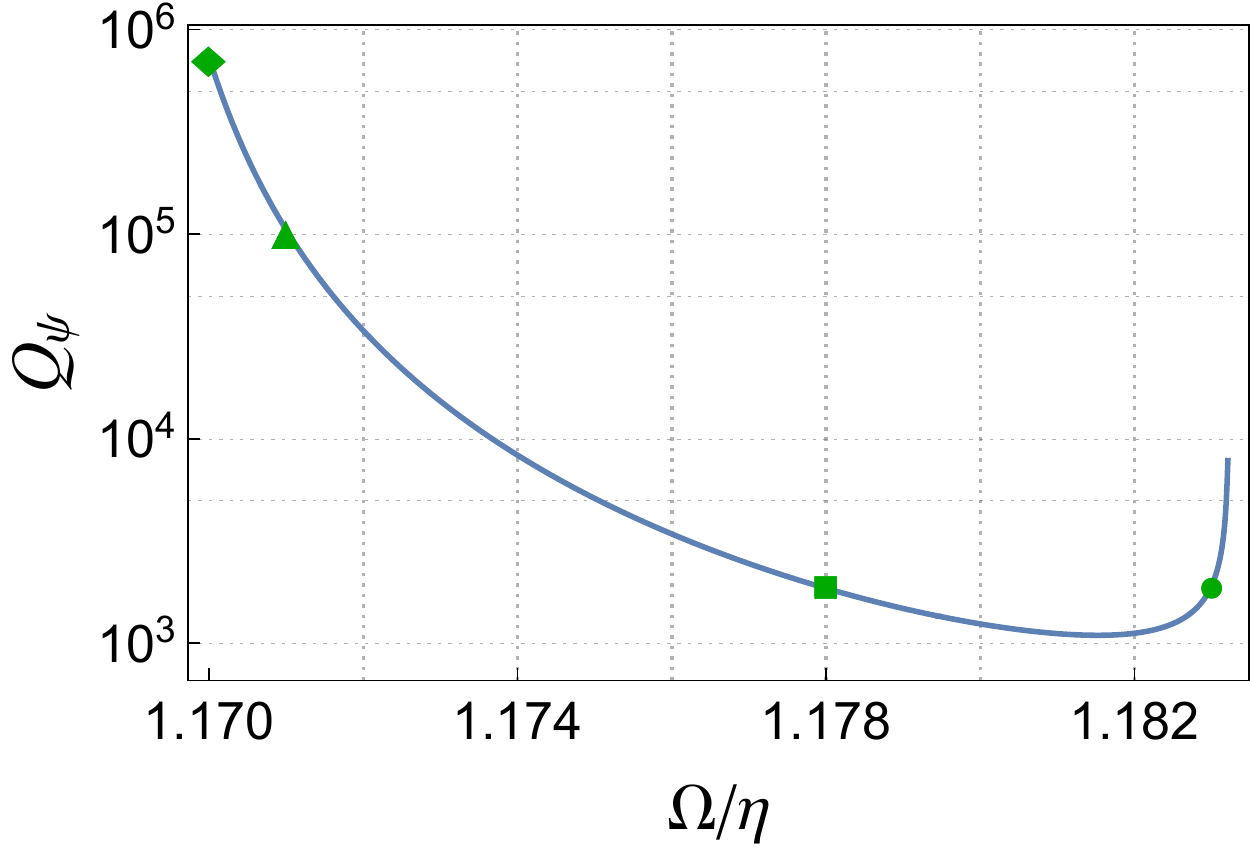}
\\
\caption{
The total charge of $\psi$, $Q_\psi$, is plotted as a function of $\Omega$.
$Q_\psi$ diverges at $\Omega=\Omega_\text{min}$ and $\Omega=\Omega_\text{max}$.
The circle, square, triangle, and diamond marks in the figure correspond to the cases of
$\Omega=1.183, 1.178, 1.171$, and $1.170$ that are shown
in  Fig.\ref{fig:configurations_1p4} and Fig.\ref{fig:charge_mu1p4}, respectively.
\label{fig:total_charge}
}
\end{figure}

As the parameter $\Omega$ varies,
the total charge of $\psi$, $Q_\psi$, defined by \eqref{eq:Q_psi} varies
as shown in Fig.\ref{fig:total_charge}.
The solution exists for $\Omega$ in the range
\begin{align}
	\Omega_\text{min}<\Omega<\Omega_\text{max},
\label{eq:Omega_range}
\end{align}
where the values of $\Omega_\text{min}$ and $\Omega_\text{max}$ are discussed later.
As seen in Fig.\ref{fig:total_charge},
$Q_\psi$ diverges at $\Omega=\Omega_\text{min}$ and $\Omega=\Omega_\text{max}$.
For $\Omega$ near $\Omega_\text{min}$ in the range \eqref{eq:Omega_range},
the solutions represent homogeneous balls,
where the radius of the ball increases as $\Omega$ approaches to $\Omega_\text{min}$,
while the constant values of $u$, $f$ and $\alpha$ are independent of $\Omega$.

%%%%%%%%%%%%%%%%%%%%%%%%%%%%%%%%%%%%%%%%%%%%%%%%%%%%%%%%%%%%%%%%%%%%%%%%%%%%%%%%%%%%%%
%%%%%%%%%%%%%%%%%%%%%%%%%%%%%%%%%%%%%%%%%%%%%%%%%%%%%%%%%%%%%%%%%%%%%%%%%%%%%%%%%%%%%%

Here, we estimate the value of $\Omega_\text{max}$.
Since $u$ is small at a large distance, and $f-\eta$ and $\alpha$ are smaller
than $u$ there (see Fig.\ref{fig:configurations_1p4}),
then solving the linearized equations of \eqref{eq:eq_u}, we have
\begin{align}
  	&u(r)\propto  \frac{1}{r}\exp \left(-\sqrt{m_{\psi}^2-\Omega^2}~~r\right).
 \label{eq:u asymptotic inf}
\end{align}
If we require the solutions are localized in a finite region,
the parameter $\Omega$ should satisfies
\begin{align}
  \Omega^2 < \Omega_\text{max}^2:=m_{\psi}^2=\mu \eta^2.
 \label{eq:Omega_max}
\end{align}

%%%%%%%%%%%%%%%%%%%%%%%%%%%%%%%%%%%%%%%%%%%
%--------------------------------------------------------
\section{Homogeneous ball solutions}
%--------------------------------------------------------
%%%%%%%%%%%%%%%%%%%%%%%%%%%%%%%%%%%%%%%%%%%
For the parameter $\Omega$ very closed to $\Omega_\text{min}$,
the homogeneous ball solutions with large radius appear.
We inspect the homogeneous ball solutions in detail.

The set of equations \eqref{eq:eq_u}, \eqref{eq:eq_f}, and \eqref{eq:eq_alpha} can be derived
from the effective action in the form
\begin{align}
	&S_\text{eff} = \int  r^2 dr \left(
		 \biggl(\frac{du}{dr}\biggr)^2 + \biggl(\frac{df}{dr}\biggr)^2 -\frac12 \biggl(\frac{d\alpha}{dr}\biggr)^2
		- U_\text{eff}(u, f,\alpha)	\right),
\label{eq:S_eff}
\\
	&U_\text{eff}(u, f,\alpha) := -\frac{\lambda}{4}(f^2-\eta^2)^2-\mu f^2 u^2
		+e^2f^2 \alpha^2 +(e\alpha-\Omega)^2u^2 .
\end{align}
If we regard the coordinate $r$ as a \lq time\rq, the effective
action \eqref{eq:S_eff} describes a mechanical system of three degrees of freedom,
$u$, $f$ and $\alpha$, where the \lq kinetic\rq\ term of $\alpha$ has the wrong sign.
In the case of the homogeneous ball solution with a large radius, the damping terms that
proportional to $1/r$ in \eqref{eq:eq_u}, \eqref{eq:eq_f}, and \eqref{eq:eq_alpha} are negligible.
In this case,
\begin{align}
	E_\text{eff}:=\biggl(\frac{du}{dr}\biggr)^2
		+ \biggl(\frac{df}{dr}\biggr)^2 -\frac12 \biggl(\frac{d\alpha}{dr}\biggr)^2
		+ U_\text{eff}(u,f,\alpha)
\end{align}
is conserved during the motion in the fictitious time $r$.

There are stationary points of the dynamical system on which
\begin{align}
	\frac{\partial U_\text{eff}}{\partial u}=0,
\quad
	\frac{\partial U_\text{eff}}{\partial f}=0,
\quad \text{and}\quad
	\frac{\partial U_\text{eff}}{\partial \alpha}=0
\label{eq:stationary_cond}
\end{align}
are satisfied.
Two stationary points exist in the region $u\leq 0, f\leq 0$, and $\alpha\leq 0$.
One stationary point, say ${\rm P_v}$, exists at $(u, f, \alpha)=(0, \eta, 0)$,
that is the true vacuum.
The other stationary point, say ${\rm P_0}$, exists at $(u, f, \alpha)=(u_0, f_0, \alpha_0)$,
where $u_0, f_0$, and $\alpha_0$ are given by solving \eqref{eq:stationary_cond} as
\begin{equation}
\begin{split}
	&\alpha_0 =\frac{1}{e(4\mu-\lambda)}
		\left((\mu-\lambda)\Omega+\sqrt{\mu(2\lambda+\mu)\Omega^2
		-\mu\lambda(4\mu-\lambda)\eta^2}\right),
\\
	&f_0 = \frac{1}{\sqrt{\mu}}(\Omega -e\alpha_0),
\\
 	&u_0 = \frac{1}{\sqrt{\mu}}\sqrt{e\alpha_0(\Omega -e\alpha_0)}.
\end{split}
  \label{eq:central_values}
\end{equation}
We note that $0 < e \alpha_0 <\Omega$ should hold for real value of $u_0$.
This condition with \eqref{eq:Omega_max} requires
\begin{align}
	\lambda < \mu .
\label{eq:lambda_mu}
\end{align}

\begin{figure}[!ht]
\centering
\includegraphics[width=10cm]{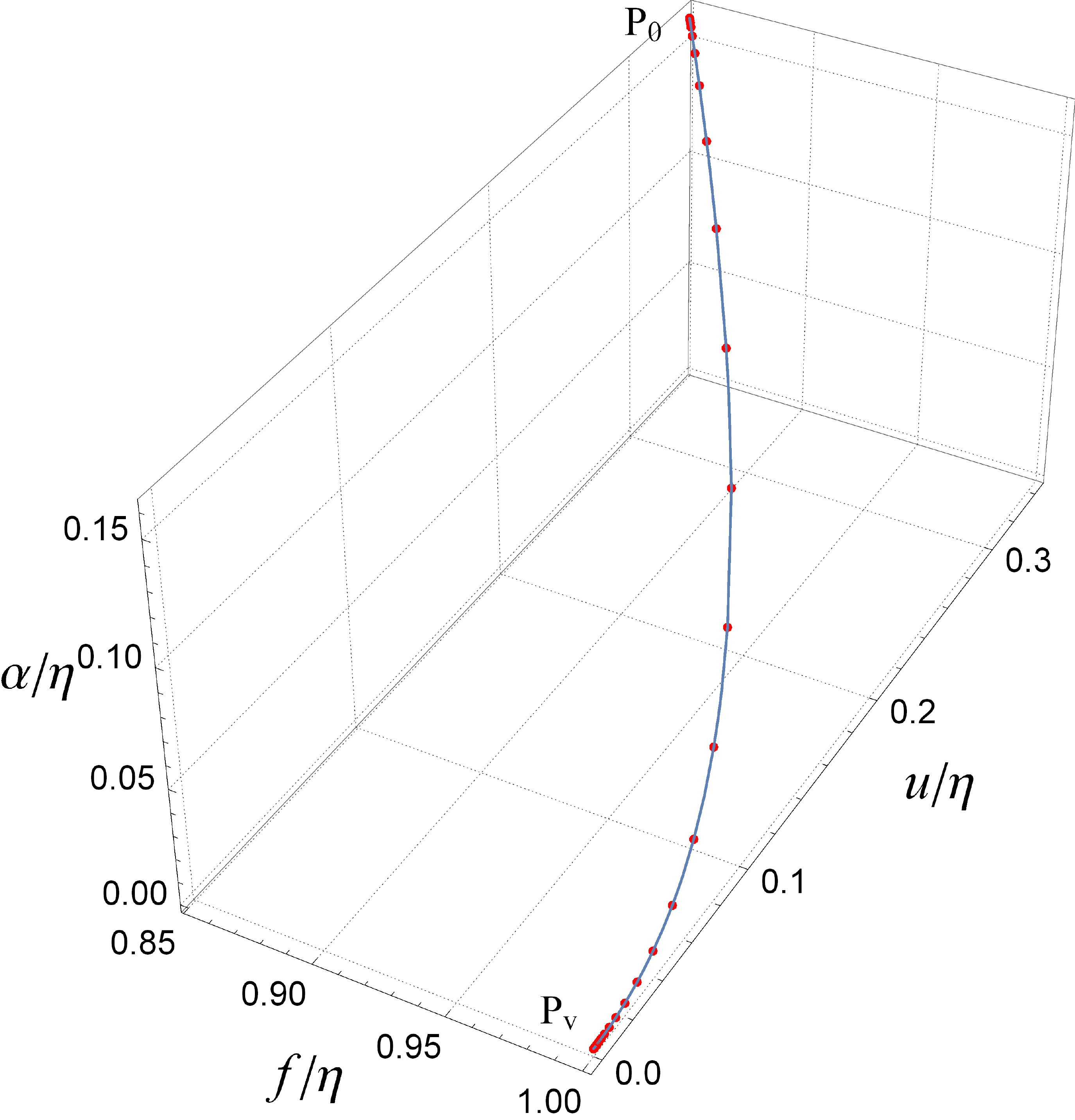}
\caption{
Trajectory of the numerical solution for $\Omega=1.170$ in the $(u, f, \alpha)$
space. It starts from a point in a vicinity of ${\rm P_0}$ and ends at ${\rm P_v}$.
Dots on the trajectory denote laps of the fictitious time $r$.
}
\label{fig:trajectory}
\end{figure}

A homogeneous ball solution with a large radius is described by a bounce solution
from ${\rm P_0}$ to ${\rm P_v}$.
Consider a point in the three-dimensional space $(u, f, \alpha)$ whose motion is governed by
equations of motion \eqref{eq:eq_u}, \eqref{eq:eq_f}, and \eqref{eq:eq_alpha}.
The point that starts in the vicinity of the stationary point ${\rm P_0}$
spends much \lq time\rq, $r$, near ${\rm P_0}$, and traverses to
the stationary point ${\rm P_v}$ in a short period, and finally stays on ${\rm P_v}$.
In Fig.\ref{fig:trajectory}, the homogeneous ball solution for $\Omega=1.170$
is shown as a trajectory in the $(u, f, \alpha)$ space.

%%%%%%%%%%%%%%%%%%%%%%%%%%%%%%%%%%%%%%%%%%%%%%%%%%%%%%%%%%%%%%%%%%%%%%%%%%%%%
%%%%%%%%%%%%%%%%%%%%%%%%%%%%%%%%%%%%%%%%%%%%%%%%%%%%%%%%%%%%%%%%%%%%%%%%%%%%%

If $\Omega$ approaches to $\Omega_\text{min}$,
the radius of the homogeneous ball diverges.
It means that the solution with infinitely large radius starts from ${\rm P_0}$.
Since $E_\text{eff}$ is conserved for the homogeneous ball solution with a large radius,
the bounce solution that describes the homogeneous ball connects the two
stationary points with equal potential heights, i.e.,
\begin{align}
	U_\text{eff}({\rm P_v} )=U_\text{eff}({\rm P_0} ).
\end{align}
We see that this occurs for
\begin{align}
  \Omega = \Omega_\text{min} := \sqrt{2\sqrt{\lambda\mu}-\lambda}~ \eta
		=\sqrt{m_{\phi}(2m_{\psi}-m_{\phi})}.
\label{eq:Omega_min}
\end{align}
Then, for the parameters satisfying \eqref{eq:lambda_mu}, we see
\begin{align}
  \Omega_\text{min} <\Omega_\text{max}.
\label{eq:Omega_min_max}
\end{align}
Then, the non-topological soliton solutions exist for the model parameters with \eqref{eq:lambda_mu}.

\if0 %%%%%%%%%%%%%%%%%%%%%%%%%%%%%%%%%%%%%%%%%%%%%%%%%%%%%%%%%%
the allowed range of $\Omega$, \eqref{eq:Omega_range} is rewritten as
\begin{align}
	2m_{\psi}m_{\phi}-m_{\phi}^2 < \Omega^2 < m_\psi^2 ,
\end{align}
equivalently,
\begin{align}
	2\sqrt{\lambda\mu}-\lambda < \left(\frac{\Omega}{\eta}\right)^2 < \mu.
\end{align}
\fi %%%%%%%%%%%%%%%%%%%%%%%%%%%%%%%%%%%%%%%%%%%%%%%%%%%%%%%%%%%%%%

%--------------------------------
%--------------------------------

\begin{figure}[!h]
\centering
\includegraphics[width=7.3cm]{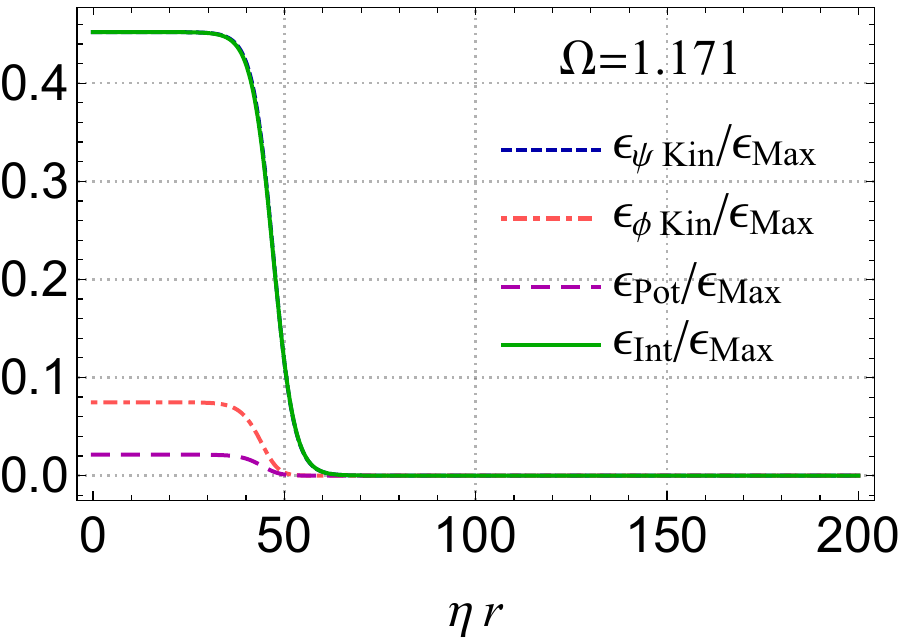}~~
\includegraphics[width=7.3cm]{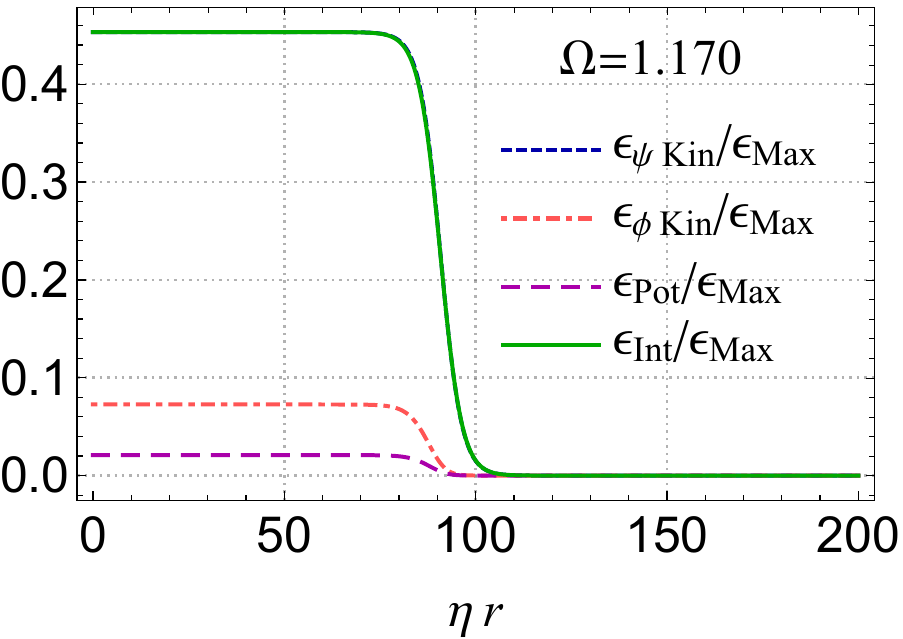}
\caption{
Components of energy densities of the homogeneous balls normalized by the central value of total energy
density are drawn for $\Omega=1.171$ (left panel)
and for $\Omega=1.170$ (right panel).
\label{fig:energy_1p4}
}
\end{figure}

Using the ansatz \eqref{eq:f}, \eqref{eq:u}, and \eqref{eq:alpha},
we rewrite the energy \eqref{eq:energy} for the symmetric system as
\begin{align}
	&E_\text{NTS}=4\pi \int_0^{\infty}r^2\epsilon(r)dr,
\label{eq:energy2}
\\
	&\epsilon := \epsilon_\text{$\psi$Kin}+\epsilon_\text{$\phi$Kin}
		+\epsilon_\text{$\psi$Elast}+\epsilon_\text{$\phi$Elast}
		+\epsilon_\text{Int} +\epsilon_\text{Pot}+\epsilon_\text{ES},
\label{eq:total_energy_density}
\end{align}
where
\begin{align}
	&\epsilon_\text{$\psi$Kin}:=\left|D_{t}\psi\right|^2
		=(e\alpha-\Omega)^2u^2, \quad
	\epsilon_\text{$\phi$Kin}:=\left|D_{t}\phi\right|^2=e^2f^2\alpha^2,
\cr
	&\epsilon_\text{$\psi$Elast}:=(D_{i}\phi)^{\ast}(D^{i}\psi)=\biggl(\frac{du}{dr}\biggr)^2,\quad
	\epsilon_\text{$\phi$Elast}:=(D_{i}\phi)^{\ast}(D^{i}\phi)=\biggl(\frac{df}{dr}\biggr)^2, \quad
\cr
	&\epsilon_\text{Pot}:=V(\phi)=\frac{\lambda}{4}(f^2-\eta)^2, \quad
	\epsilon_\text{Int}:=\mu|\phi|^2|\psi|^2=\mu f^2u^2, \quad
	\epsilon_\text{ES}:=\frac{1}{2}E_iE^i=\frac{1}{2}\biggl(\frac{d\alpha}{dr}\biggr)^2,
\label{eq:energy_density}
\end{align}
are densities of kinetic energy of $\psi$ and $\phi$, elastic energy of $\psi$ and $\phi$,
potential energy of $\phi$, interaction energy between $\psi$ and $\phi$,
and electrostatic energy, respectively.
For the homogeneous ball solutions,
these components of energy density are shown in Fig.\ref{fig:energy_1p4}.
The dominant components of the energy density $\epsilon$ are $\epsilon_\text{$\psi$Kin}$
and $\epsilon_\text{Int}$, and subdominant components are $\epsilon_\text{$\phi$Kin}$
and $\epsilon_\text{Pot}$ for the present cases.
The densities of the elastic energy and the electrostatic energy,
which appear near the surface of the ball,
are negligibly small, then, they are not plotted.

We see, from \eqref{eq:central_values}, that the dominant and subdominant components of energy density
inside the balls are constants with the values
\begin{align}
	&\epsilon_\text{$\psi$Kin}=\epsilon_\text{Int}=\frac{1}{\mu} e\alpha_0(\Omega-e\alpha_0)^3,
\cr
	&\epsilon_\text{$\phi$Kin}=\frac{1}{\mu} (e\alpha_0)^2(\Omega-e\alpha_0)^2, \quad
	\epsilon_\text{Pot}=\frac{\lambda}{\mu^2} \left((\Omega-e\alpha_0)^2-\eta^2\right)^2.
\end{align}
Then the energy density and pressure\footnote{See Appendix \ref{E_M_tensor}.}
for the homogeneous ball are constants given by
\begin{align}
	\epsilon & \simeq \epsilon_\text{$\psi$Kin}+\epsilon_\text{$\phi$Kin}
		+\epsilon_\text{Int}+\epsilon_\text{Pot}
\cr
		&=\frac{2}{\mu} e\alpha_0(\Omega-e\alpha_0)^3 +\frac{1}{\mu} (e\alpha_0)^2(\Omega-e\alpha_0)^2
			+\frac{\lambda}{\mu^2} \left((\Omega-e\alpha_0)^2-\eta^2\right)^2,
\cr
	p &=p_r \simeq p_\theta = p_\varphi
		\simeq \epsilon_\text{$\psi$Kin}+\epsilon_\text{$\phi$Kin}
		-\epsilon_\text{Int}-\epsilon_\text{Pot}
\cr
		&=\frac{1}{\mu} (e\alpha_0)^2(\Omega-e\alpha_0)^2
			-\frac{\lambda}{\mu^2} \left((\Omega-e\alpha_0)^2-\eta^2\right)^2.
\end{align}
We see that the pressure is almost isotropic,
and $p\sim 0.05 \epsilon$ for the homogeneous ball of $\Omega=1.170$.
The equation of state of the homogeneous balls is like non-relativistic gas.

%%%%%%%%%%%%%%%%%%%%%%%%%%%%%%%%%%%%%%%%%%%%%%%%%%%%%%%%%%%%%%%%%%%%%%%%%%%%%%%%%
%%%%%%%%%%%%%%%%%%%%%%%%%%%%%%%%%%%%%%%%%%%%%%%%%%%%%%%%%%%%%%%%%%%%%%%%%%%%%%%%%

In the limit $\Omega \to \Omega_\text{min}$, so that $Q_\psi \to \infty$, we see
\begin{align}
\if0 %%%%
	&e \alpha_0= \frac{\lambda^{1/4} (\sqrt{\mu}-\sqrt{\lambda})}{\sqrt{2\sqrt{\mu}-\sqrt{\lambda}}}\eta,
	\quad
	\Omega_\text{min}-e \alpha_0= \frac{\lambda^{1/4} \sqrt{\mu}}{\sqrt{2\sqrt{\mu}-\sqrt{\lambda}}} \eta
\cr
\fi %%%%
	&\epsilon_\text{$\psi$Kin}=\epsilon_\text{Int}
	\to \frac{\lambda (\sqrt{\mu}-\sqrt{\lambda})\sqrt{\mu}}{(2\sqrt{\mu}-\sqrt{\lambda})^2} \eta^4,
\cr
	&\epsilon_\text{$\phi$Kin}
	\to \lambda\left(\frac{\sqrt{\mu}-\sqrt{\lambda}}{2\sqrt{\mu}-\sqrt{\lambda}}\right)^2 \eta^4,
\quad
	\epsilon_\text{Pot}
	\to \lambda\left(\frac{\sqrt{\mu}-\sqrt{\lambda}}{2\sqrt{\mu}-\sqrt{\lambda}}\right)^2 \eta^4,
\end{align}
then we have
\begin{align}
&\epsilon
	\to \frac{2\lambda (\sqrt{\mu}-\sqrt{\lambda})}{2\sqrt{\mu}-\sqrt{\lambda}} \eta^4,
\cr
	&p \to 0.
\label{eq:energy_density_limit}
\end{align}
Therefore, in the large homogeneous ball limit, the  ball becomes dust ball with constant energy density
given by \eqref{eq:energy_density_limit}.

%%%%%%%%%%%%%%%%%%%%%%%%%%%%%%%%%%%%%%%%%%%
\section{Stability}
%%%%%%%%%%%%%%%%%%%%%%%%%%%%%%%%%%%%%%%%%%%

The nontopological soliton, called Q-ball, can be interpreted as a condensate of particles of
the scalar field $\psi$,
where the Higgs field plays the role of glue against repulsive force
by the $U(1)$ gauge field.
We compare energy of the soliton, $E_\text{NTS}$, given by \eqref{eq:energy2}
with mass energy of the free particles of $\psi$
that have the same amount of charge of the soliton as a whole.
Then, the numbers of the particles is defined by
\begin{align}
  	N_\psi &:=\frac{Q_{\psi}}{e},
  \label{eq:numbers_psi}
\end{align}
and the mass energy of the free particles of $\psi$ is given by
$E_\text{free} = m_{\psi} N_{\psi}$.

\begin{figure}[!h]
\centering
\includegraphics[width=8cm]{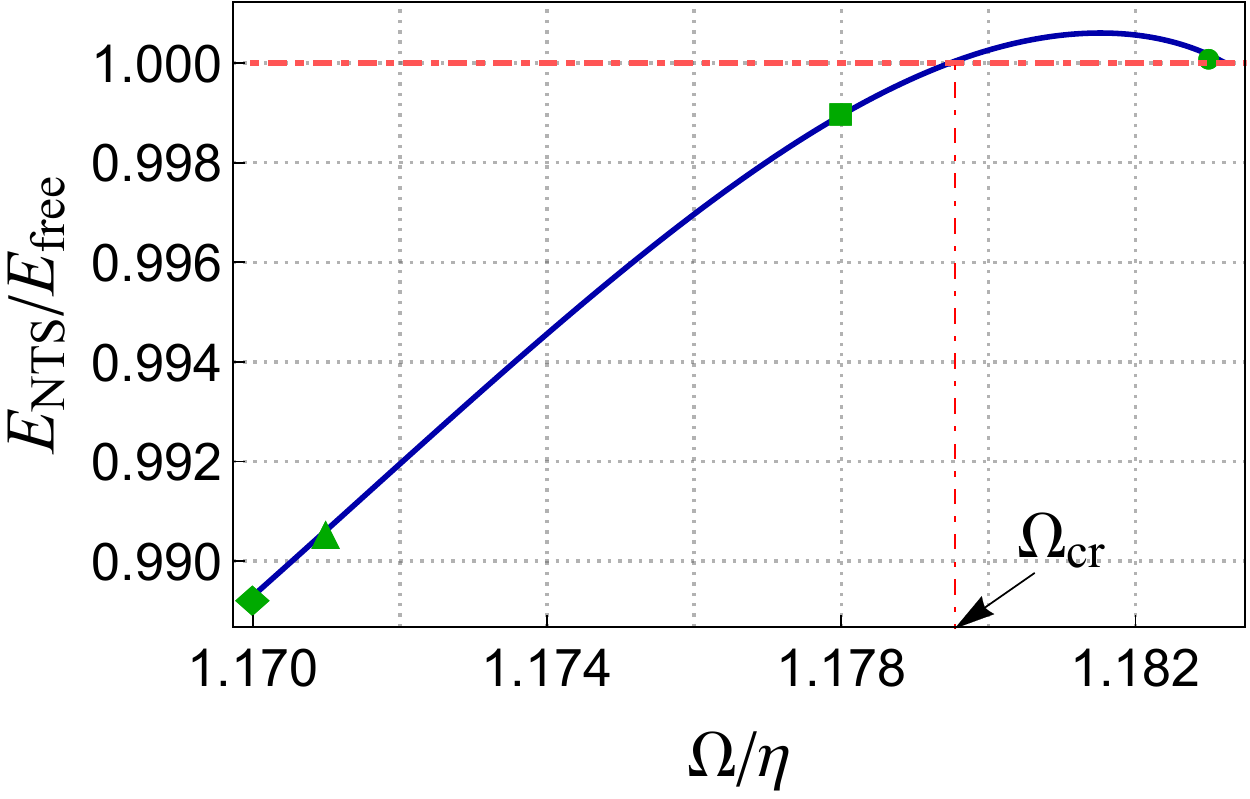}~~
\centering
\includegraphics[width=8cm]{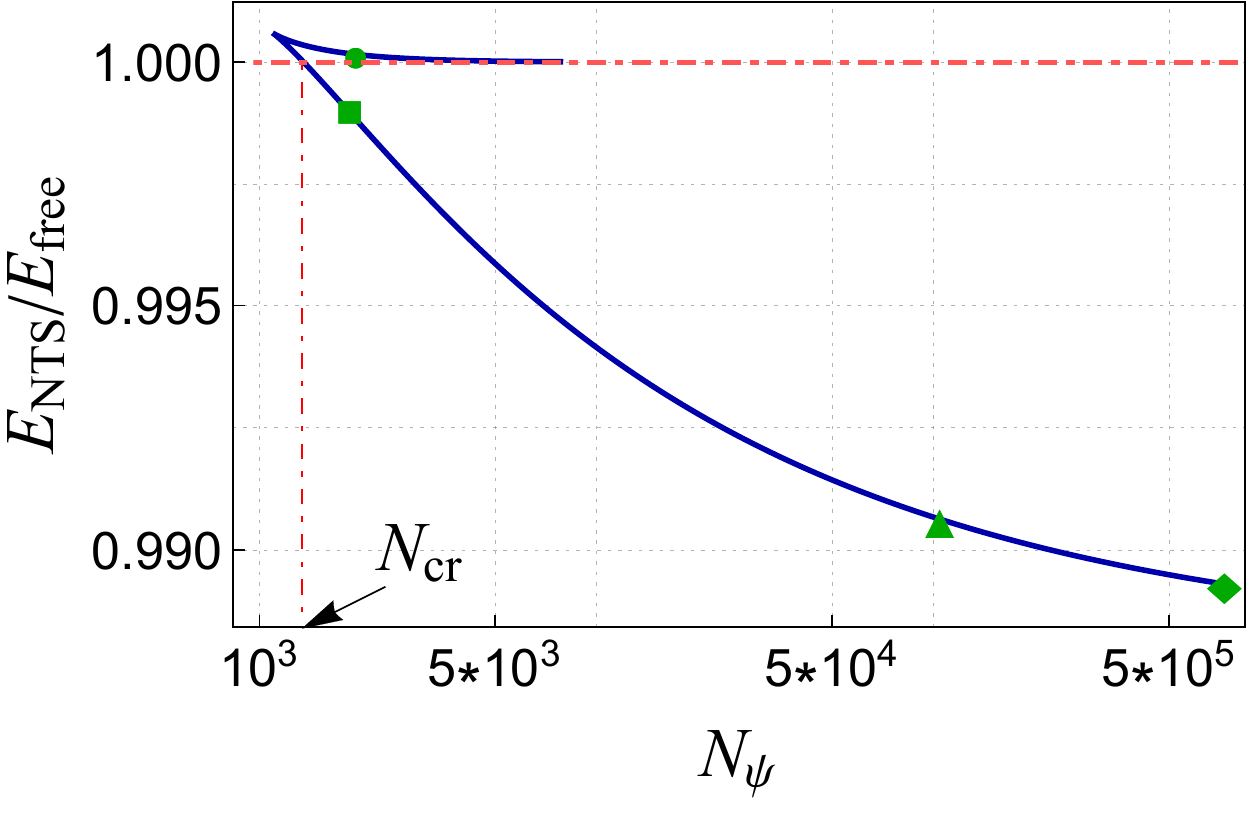}
\caption{
The energy ratio $E_\text{NTS}/E_\text{free}$ is plotted as a function of $\Omega$ (left panel),
and as a function of $N_\psi$ (right panel).
The circle, square, triangle, and diamond marks in the figure correspond to the cases of
$\Omega=1.183, 1.178, 1.171$, and $1.170$ that are shown
in  Figs.\ref{fig:configurations_1p4} and \ref{fig:charge_mu1p4}, respectively.
\label{fig:energy ratio}
}
\end{figure}

Fig.\ref{fig:energy ratio} shows the energy ratio $E_\text{NTS}/E_\text{free}$
as a function of $\Omega$ and as a function of $N_\psi$, respectively.
We find a critical value of $\Omega$, $\Omega_\text{cr}$, such that if
$\Omega<\Omega_\text{cr}$, $E_\text{NTS}<E_\text{free}$ holds.
Therefore, a Q-ball for $\Omega$ in the range
\begin{align}
  \Omega_\text{min}<\Omega<\Omega_\text{cr}
\label{eq:NTS_condition}
\end{align}
is energetically preferable than the free $\psi$ particles with the same charge of the Q-ball as a whole.
From the Fig.\ref{fig:energy ratio},
there exist stable Q-balls that are condensates of large numbers of $\psi$ particles.

Since the energy density and charge density are constant inside the ball,
the total energy and the total charge of matter field of the homogeneous ball
are written by
\begin{align}
	E_\text{NTS} = \epsilon V, \quad\text{and}\quad
	Q_\psi = \rho_\psi V,
\end{align}
where $V$ is the volume of the ball.
Then, the energy ratio $E_\text{NTS}/E_\text{free}$ for the homogeneous ball is
calculated as
\begin{align}
	\frac{E_\text{NTS}}{E_\text{free}}
	=\frac{\epsilon V}{m_\psi N_\psi}= \frac{\epsilon Q_\psi/\rho_\psi}{m_\psi Q_\psi/e}
	= \frac{e \epsilon }{m_\psi \rho_\psi}.
\end{align}
In the limit $\Omega \to \Omega_\text{min}$, so that $Q_\psi \to \infty$,
we obtain $E_\text{NTS}/E_\text{free}$
as
\begin{align}
	\frac{E_\text{NTS}}{E_\text{free}}
	\to \left(\left(2-\sqrt{\lambda/\mu}\right)\sqrt{\lambda/\mu}\right)^{1/2}.
\end{align}
It is clear that ${E_\text{NTS}}/{E_\text{free}}<1$ for $ \lambda <\mu $
in the limit $\Omega \to \Omega_\text{min}$.
Therefore, in the large limit of the homogeneous ball solution is stable.

%%%%%%%%%%%%%%%%%%%%%%%%%%%%%%%%%%%%%%%%%%%%%%%%%%%%%%%%%%%%%%%%%%%%%%%%
%%%%%%%%%%%%%%%%%%%%%%%%%%%%%%%%%%%%%%%%%%%%%%%%%%%%%%%%%%%%%%%%%%%%%%%%

We show $E_\text{NTS}/E_\text{free}$ for various $Q_\psi$ in Table.\ref{tab:E_NTS}.
We see the inequality
\begin{align}
	E_\text{NTS}(Q_{\psi 1}) + E_\text{NTS}(Q_{\psi 2})
		>	E_\text{NTS}(Q_{\psi 1}+Q_{\psi 2})
\end{align}
holds for any $Q_{\psi 1}$ and $Q_{\psi 2}$ in the table.
It means that one large Q-ball is energetically preferable to two small Q-balls.
Therefore, two Q-balls can merge into a Q-ball, but a Q-ball does not decay into two Q-balls.

%%%%%%%%%%%%%%%%%%%%%%%%%%%%%%%%%%%%%%%%%%%%%%%%%%%%%%%%%%%%%%%%%
\begin{table}[!ht]
\centering
\begin{tabular}{c|r|c}
\hline \hline
  $\Omega$ &$Q_{\psi}~~$ & $E_\text{NTS}$ \\
\hline \hline
  1.17771 & 2000 & 2363.4 \\ \hline
  1.17559 & 4000 & 4716.3 \\ \hline
  1.17465 & 6000 & 7066.4 \\ \hline
  1.17407 & 8000 & 9415.1 \\ \hline
  1.17368 & 10000 & 11762.8 \\ \hline
  1.17262 & 20000 & 23493.5 \\ \hline
  1.17213 & 30000 & 35217.0 \\ \hline
  1.17182 & 40000 & 46936.7 \\ \hline
  1.17161 & 50000 & 58653.7 \\ \hline
  1.17103 & 100000 & 117217 \\ \hline
  1.17059 & 200000 & 234295 \\ \hline
  1.17037 & 300000 & 351342 \\ \hline
  1.17024 & 400000 & 468373 \\ \hline
  1.17015 & 500000 & 585392 \\ \hline
\hline
\end{tabular}
\caption{
The total charge of $\psi$, $Q_\psi$, and total energy, $E_\text{NTS}$,
of Q-balls for various values of parameters $\Omega$.
}
\label{tab:E_NTS}
\end{table}
%%%%%%%%%%%%%%%%%%%%%%%%%%%%%%%%%%%%%%%%%%%%%%%%%%%%%%%%%%%%%%%%%

%%%%%%%%%%%%%%%%%%%%%%%%%%%%
\section{Summary and discussions}
%%%%%%%%%%%%%%%%%%%%%%%%%%%%

In this paper, we have studied the coupled system of a complex matter scalar field,
a U(1) gauge field, and a complex Higgs scalar field with a potential
that causes spontaneous symmetry breaking.
This is a generalization of the Friedberg-Lee-Sirlin model \cite{Friedberg:1976me}.
In this system, a local U(1)$\times$ global U(1) symmetry is broken spontaneously
into a global U(1) symmetry by the Higgs field.
We have shown numerically that there are spherically symmetric
nontopological soliton solutions, Q-balls, that are characterized
by phase rotation of the complex matter scalar field, $\Omega$.
The Q-balls can exists for a finite range of $\Omega$,
and there are two types of solutions: Gaussian balls and homogeneous balls.

In the homogeneous ball solutions, the fields take constant values inside the ball,
and they change the values quickly at the ball surface to the vacuum values outside the ball.
The charge density of matter scalar field that arises inside the ball is canceled out everywhere
by the counter charge cloud of the Higgs and the gauge fields,
namely, perfect screening occurs \cite{Ishihara:2018eah, Ishihara:2018rxg}.
Inspecting the energy-momentum tensor of the fields, we have shown that energy density and pressure
inside the balls take constant values. The pressure is almost isotropic,
and the value is much less than the energy density.
Then, a homogeneous ball is like a ball of homogeneous nonrelativistic gas.

Homogeneous ball solutions appear as \lq Q-matters\rq\
in the system of a self-coupling single complex scalar field
studied by Coleman \cite{Coleman:1985ki}. These solutions are interpreted as bounce
solutions that connect two stationary points of the potential of one degree of freedom.
In the extended system by introducing a U(1) gauge field,
the homogeneous ball solution does not appear.
In the gauged system, since repulsive force acting between charges pushes them outward
to the surface of the ball, then the solution has
radial inhomogeneity \cite{Lee:1988ag,Shi:1991gh,Gulamov:2015fya}.
In contrast, in the gauged system with spontaneous symmetry breaking investigated in this paper,
the perfect screening of charge occurs, then no repulsive force acts inside the ball.
Therefore, the homogeneous ball solutions can exist.
This is suggested in the work \cite{Anagnostopoulos:2001dh}.
The homogeneous ball solutions are interpreted as bounce solutions
that connect two stationary points of the potential of three degrees of freedom.
Then, the homogeneous ball solutions obtained in this paper are extensions of Coleman's Q-matters.

By comparison of the energies,
it was shown that if the charge of the matter field
is greater than a critical value, a Q-ball is stable against dispersion
into free particles and against decay into two smaller Q-balls.
In addition to the analysis in this paper, it is important to investigate the stability
in various view points \cite{Cohen:1986ct, Kusenko:1997ad, Multamaki:1999an, Paccetti:2001uh,
Kawasaki:2005xc, Sakai:2007ft}.

In the extended systems by the gauge field without the Higgs field, the size of a stable charged
Q-ball has upper bound\cite{Lee:1988ag,Shi:1991gh, Gulamov:2015fya}.
In contrast, a stable charge screened homogeneous ball has no limit of mass.
Of cause, this is true as far as the gravity can be neglected.
If the mass of the homogeneous ball becomes too large so that the pressure fails to sustain the gravity,
the ball would collapse to a black hole. Then, there exists upper bound of mass for the stable
homogeneous ball if the gravity is taken into account. It is an interesting issue to study
the gravitational effects
on the Q-balls \cite{Friedberg:1986tp, Friedberg:1986tq, Lee:1986tr, Lynn:1988rb, Mielke:2002bp}.
We would report this issue on the present system in a forthcoming paper.

The Q-balls obtained in this paper would have applications in cosmology and
astrophysics. The perfect screening of the charge is a preferable property for the gauged
Q-balls to be dark matter \cite{Kusenko:1997si, %Kawasaki:2008zz,
Kusenko:2001vu, %Multamaki:2002hv,
Fujii:2001xp, Enqvist:2001jd, Kusenko:2004yw}.
It is an important issue how much amount of the Q-balls are produced
in the evolution of the universe \cite{Frieman:1988ut, Griest:1989bq,  Kasuya:2000wx,
Postma:2001ea, Multamaki:2002hv, Hiramatsu:2010dx}.
It would be an interesting problem
to clarify the mass distribution spectrum of the Q-balls, which would evolve by merging process of Q-balls,
in the present stage of the universe.

In the model studied in this paper, we assumed that matter is described by a complex scalar
field, for simplicity. It is interesting to consider fermionic matter fields that form Q-balls.
Indeed, fermionic Q-balls are already studied \cite{ %Lee:1988ag,
Friedberg:1976eg, Friedberg:1977xf, Shima:1977tm, Levi:2001aw},
but a large fermionic soliton is hardly produced because of the Pauli exclusion principle.
If two fermions make a bosonic bound state as in a superconductor,
it is expected that the charge screened large Q-ball as was discussed in this paper,
would be possible. To clarify this possibility would be a challenging work.

%%%%%%%%%%%%%%%%%%%%%%%%%%%%%%%%%%%%%%%%%%%
\section*{Acknowledgements}
%%%%%%%%%%%%%%%%%%%%%%%%%%%%%%%%%%%%%%%%%%%

We would like to thank K.-i. Nakao, H. Itoyama, Y. Yasui, N. Maru, N. Sakai, and M. Minamitsuji
for valuable discussion.
H.I. was supported by JSPS KAKENHI Grant Number 16K05358.

%%%%%%%%%%%%%%%%%%%%%%%%%%%%%%%%%%%%%%%%%%%%%%%%%%%%%%%%%%%%%%%%
\appendix
%%%%%%%%%%%%%%%%%%%%%%%%%%%%%%%%%%%%%%%%%%%%%%%%%%%%%
\section{Energy-Momentum Tensor of the System}
\label{E_M_tensor}
%%%%%%%%%%%%%%%%%%%%%%%%%%%%%%%%%%%%%%%%%%%%%%%%%%%%%
The energy-momentum tensor $T_{\mu\nu}$ of the present system is given by
\begin{align}
T_{\mu\nu}
	=&2(D_{\mu}\psi)^{\ast}(D_{\nu}\psi)
	-g_{\mu\nu}
		(D_{\alpha}\psi)^{\ast}(D^{\alpha}\psi)
\cr
	&+2(D_{\mu}\phi)^{\ast}(D_{\nu}\phi)
	-g_{\mu\nu}(D_{\alpha}\phi)^{\ast}(D^{\alpha}\phi)
\cr
	&-g_{\mu\nu}\left( V(\phi)
	 + \mu \psi^{\ast}\psi \phi^{\ast} \phi \right)
\cr
	&+\left( F_{\mu\alpha}F_{\nu}^{~\alpha}
	-\frac{1}{4}g_{\mu\nu}F_{\alpha\beta}F^{\alpha\beta}\right).
\label{eq:T_munu}
\end{align}
Energy density and pressure components are given by
\begin{align}
 	\epsilon=&-T_t^t
\cr
		= &\left|D_{t}\psi\right|^2  +(D_{i}\psi)^{\ast}(D^{i}\psi)
		 +\left|D_{t}\phi\right|^2+(D_{i}\phi)^{\ast}(D^{i}\phi)
\cr
	& +V(\phi)+\mu|\psi|^2|\phi|^2 +\frac{1}{2}\left(E_iE^i+B_iB^i\right),
\label{eq:T_tt}
\end{align}
\vspace{-15mm}
\begin{align}
 	p_r=&T_r^r
\cr		= &(D_r\psi)^{\ast}(D^r \psi)+\left|D_{t}\psi\right|^2
			-(D_{\theta}\psi)^{\ast}(D^{\theta}\psi) -(D_{\varphi}\psi)^{\ast}(D^{\varphi}\psi)
\cr
		 &+(D_r\phi)^{\ast}(D^r \phi)+\left|D_{t}\phi\right|^2
			-(D_{\theta}\phi)^{\ast}(D^{\theta}\phi)-(D_{\varphi}\phi)^{\ast}(D^{\varphi}\phi)
\cr
	  	& -V(\phi) -\mu|\psi|^2|\phi|^2
\cr
		&+\frac{1}{2}(-E_rE^r+E_\theta E^\theta+E_\varphi E^\varphi
		-B_r B^r+B_\theta B^\theta+B_\varphi B^\varphi),
\label{eq:T_rr}
\end{align}
\vspace{-15mm}
\begin{align}
 	p_{\theta}=&T_\theta^\theta
\cr		= &(D_\theta\psi)^{\ast}(D^\theta \psi)
		+\left|D_{t}\psi\right|^2 -(D_r\psi)^{\ast}(D^r\psi)
		-(D_{\varphi}\psi)^{\ast}(D^{\varphi}\psi)
\cr
		 &+(D_{\theta}\phi)^{\ast}(D^{\theta}\phi)+\left|D_{t}\phi\right|^2
		-(D_r\phi)^{\ast}(D^r \phi) -(D_{\varphi}\phi)^{\ast}(D^{\varphi}\phi)
\cr
	  	& -V(\phi) -\mu|\psi|^2|\phi|^2
\cr
		&+\frac{1}{2}(-E_\theta E^\theta +E_rE^r+E_\varphi E^\varphi
		-B_\theta B^\theta+B_r B^r+B_\varphi B^\varphi),
\label{eq:T_thetatheta}
\end{align}
\vspace{-15mm}
\begin{align}
 	p_{\varphi}=&T_\varphi^\varphi
\cr		= &(D_{\varphi}\psi)^{\ast}(D^{\varphi}\psi)
		+\left|D_{t}\psi\right|^2 -(D_r\psi)^{\ast}(D^r\psi)
		-(D_\theta\psi)^{\ast}(D^\theta \psi)
\cr
		 &+(D_{\varphi}\phi)^{\ast}(D^{\varphi}\phi)+\left|D_{t}\phi\right|^2
		-(D_r\phi)^{\ast}(D^r \phi) -(D_{\theta}\phi)^{\ast}(D^{\theta}\phi)
\cr
	  	& -V(\phi) -\mu|\psi|^2|\phi|^2
\cr
		&+\frac{1}{2}(-E_\varphi E^\varphi+E_rE^r+E_\theta E^\theta
		-B_\varphi B^\varphi+B_r B^r+B_\theta B^\theta).
\label{eq:T_phiphi}
\end{align}


\begin{thebibliography}{999}

\bibitem{Friedberg:1976me}
  R.~Friedberg, T.~D.~Lee and A.~Sirlin,
  Phys.\ Rev.\ D {\bf 13}, 2739 (1976).

\bibitem{Coleman:1985ki}
  S.~R.~Coleman,
  Nucl.\ Phys.\ B {\bf 262}, 263 (1985)
  Erratum: [Nucl.\ Phys.\ B {\bf 269}, 744 (1986)].

\bibitem{Kusenko:1997zq}
  A.~Kusenko,
  Phys.\ Lett.\ B {\bf 405}, 108 (1997).

\bibitem{Dvali:1997qv}
  G.~R.~Dvali, A.~Kusenko and M.~E.~Shaposhnikov,
  Phys.\ Lett.\ B {\bf 417}, 99 (1998).

\bibitem{Kasuya:2000sc}
  S.~Kasuya and M.~Kawasaki,
  Phys.\ Rev.\ Lett.\  {\bf 85}, 2677 (2000).


\bibitem{Kusenko:1997si}
A.~Kusenko and M.~E.~Shaposhnikov,
Phys.\ Lett.\ B {\bf 418}, 46 (1998).

\bibitem{Kusenko:2001vu}
  A.~Kusenko and P.~J.~Steinhardt,
  Phys.\ Rev.\ Lett.\  {\bf 87}, 141301 (2001).

\bibitem{Fujii:2001xp}
  M.~Fujii and K.~Hamaguchi,
  Phys.\ Lett.\ B {\bf 525}, 143 (2002).

\bibitem{Enqvist:2001jd}
  K.~Enqvist, A.~Jokinen, T.~Multamaki and I.~Vilja,
  Phys.\ Lett.\ B {\bf 526}, 9 (2002).

\bibitem{Kusenko:2004yw}
  A.~Kusenko, L.~Loveridge and M.~Shaposhnikov,
  Phys.\ Rev.\ D {\bf 72}, 025015 (2005).


\bibitem{Enqvist:1997si}
  K.~Enqvist and J.~McDonald,
  Phys.\ Lett.\ B {\bf 425}, 309 (1998).


\bibitem{Kasuya:1999wu}
  S.~Kasuya and M.~Kawasaki,
  Phys.\ Rev.\ D {\bf 61}, 041301 (2000).

\bibitem{Kawasaki:2002hq}
  M.~Kawasaki, F.~Takahashi and M.~Yamaguchi,
  Phys.\ Rev.\ D {\bf 66}, 043516 (2002).


\bibitem{Lee:1988ag}
  K.~M.~Lee, J.~A.~Stein-Schabes, R.~Watkins and L.~M.~Widrow,
  Phys.\ Rev.\ D {\bf 39}, 1665 (1989).

\bibitem{Shi:1991gh}
   X.~Shi and X.~Z.~Li,
   J.\ Phys.\ A {\bf 24}, 4075 (1991).
   %

\bibitem{Gulamov:2015fya}
  I.~E.~Gulamov, E.~Y.~Nugaev, A.~G.~Panin and M.~N.~Smolyakov,
  %``Some properties of U(1) gauged Q-balls,''
  Phys.\ Rev.\ D {\bf 92}, no. 4, 045011 (2015).

\bibitem{Arodz:2008nm}
  H.~Arodz and J.~Lis,
  %``Compact Q-balls and Q-shells in a scalar electrodynamics,''
  Phys.\ Rev.\ D {\bf 79}, 045002 (2009).

\bibitem{Tamaki:2014oha}
  T.~Tamaki and N.~Sakai,
  %``Large gauged Q-balls with regular potential,''
  Phys.\ Rev.\ D {\bf 90}, 085022 (2014).



\bibitem{Ishihara:2018rxg}
  H.~Ishihara and T.~Ogawa,
  arXiv:1811.10894 [hep-th], Prog. Theor. Exp. Phys. in press.

\bibitem{Ishihara:2018eah}
  H.~Ishihara and T.~Ogawa,
  arXiv:1811.10848 [hep-th].

%*****************************************
\bibitem{Anagnostopoulos:2001dh}
  K.~N.~Anagnostopoulos, M.~Axenides, E.~G.~Floratos and N.~Tetradis,
  %``Large gauged Q balls,''
  Phys.\ Rev.\ D {\bf 64}, 125006 (2001).



\bibitem{Cohen:1986ct}
  A.~G.~Cohen, S.~R.~Coleman, H.~Georgi and A.~Manohar,
  Nucl.\ Phys.\ B {\bf 272}, 301 (1986).

\bibitem{Kusenko:1997ad}
  A.~Kusenko,
  Phys.\ Lett.\ B {\bf 404}, 285 (1997).

\bibitem{Multamaki:1999an}
  T.~Multamaki and I.~Vilja,
  Nucl.\ Phys.\ B {\bf 574}, 130 (2000).

\bibitem{Paccetti:2001uh}
  F.~Paccetti Correia and M.~G.~Schmidt,
  Eur.\ Phys.\ J.\ C {\bf 21}, 181 (2001).

 \bibitem{Kawasaki:2005xc}
     M.~Kawasaki, K.~Konya and F.~Takahashi,
     %``Q-ball instability due to U(1) breaking,''
     Phys.\ Lett.\ B {\bf 619}, 233 (2005).

\bibitem{Sakai:2007ft}
    N.~Sakai and M.~Sasaki,
  Prog.\ Theor.\ Phys.\  {\bf 119}, 929 (2008).




 \bibitem{Friedberg:1986tp}
   R.~Friedberg, T.~D.~Lee and Y.~Pang,
   Phys.\ Rev.\ D {\bf 35}, 3640 (1987).

 \bibitem{Friedberg:1986tq}
   R.~Friedberg, T.~D.~Lee and Y.~Pang,
   Phys.\ Rev.\ D {\bf 35}, 3658 (1987).

\bibitem{Lee:1986tr}
  T.~D.~Lee and Y.~Pang,
  Phys.\ Rev.\ D {\bf 35}, 3678 (1987).

\bibitem{Lynn:1988rb}
  B.~W.~Lynn,
  Nucl.\ Phys.\ B {\bf 321}, 465 (1989).


\bibitem{Mielke:2002bp}
  E.~W.~Mielke and F.~E.~Schunck,
  Phys.\ Rev.\ D {\bf 66}, 023503 (2002).



\bibitem{Frieman:1988ut}
  J.~A.~Frieman, G.~B.~Gelmini, M.~Gleiser and E.~W.~Kolb,
  Phys.\ Rev.\ Lett.\  {\bf 60}, 2101 (1988).
%
\bibitem{Griest:1989bq}
  K.~Griest and E.~W.~Kolb,
  Phys.\ Rev.\ D {\bf 40}, 3231 (1989).

\bibitem{Kasuya:2000wx}
	S.~Kasuya and M.~Kawasaki,
	Phys.\ Rev.\ D {\bf 62}, 023512 (2000).
%
\bibitem{Postma:2001ea}
  M.~Postma,
  Phys.\ Rev.\ D {\bf 65}, 085035 (2002).


\bibitem{Multamaki:2002hv}
  T.~Multamaki and I.~Vilja,
  Phys.\ Lett.\ B {\bf 535}, 170 (2002).

\bibitem{Hiramatsu:2010dx}
 	T.~Hiramatsu, M.~Kawasaki and F.~Takahashi,
  	JCAP {\bf 1006}, 008 (2010).



 \bibitem{Friedberg:1976eg}
   R.~Friedberg and T.~D.~Lee,
   Phys.\ Rev.\ D {\bf 15}, 1694 (1977).

 \bibitem{Friedberg:1977xf}
   R.~Friedberg and T.~D.~Lee,
   Phys.\ Rev.\ D {\bf 16}, 1096 (1977).

 \bibitem{Shima:1977tm}
   K.~Shima,
   Nuovo Cim.\ A {\bf 44}, 163 (1978).

 \bibitem{Levi:2001aw}
   T.~S.~Levi and M.~Gleiser,
   Phys.\ Rev.\ D {\bf 66}, 087701 (2002).


\end{thebibliography}
\end{document}